\documentstyle[aps]{revtex}

\input{epsf}
\def\dofig#1#2{\epsfysize=#1 \centerline{\epsfbox{#2}}}

\begin{document}
\title{Resonance between Noise and Delay}
\author{Toru Ohira\footnote{E-mail:ohira@csl.sony.co.jp}}
\address{
Sony Computer Science Laboratory\\
3-14-13 Higashi-gotanda, Shinagawa,\\
 Tokyo 141, Japan\\
}
\author{Yuzuru Sato\footnote{E-mail:ysato@sacral.c.u-tokyo.ac.jp}}
\address{
Institute of Physics,\\
Graduate School of Arts and Science, University of Tokyo\\
3-8-1 komaba, Meguro, Tokyo 153 Japan\\
(Sony Computer Science Laboratory Technical Report: SCSL-TR-98-016)
}
\date{\today}
\maketitle
\begin{abstract}
We propose here a stochastic binary element whose transition rate
depends on its state at a fixed interval in the past.
With this delayed stochastic transition
this is one of the simplest dynamical models under the 
influence of ``noise'' and ``delay''.
We demonstrate numerically and analytically that
we can observe resonant phenomena between the oscillatory
behavior due to noise and that due to delay. 
\end{abstract}
\pacs{02.50.-r, 05.40.+j, 02.30.Ks}
\vspace{2em}

Resonant behavior is one of the most studied and utilized
fundamental physical phenomena.
As well as being of interest in a variety of fields of physics,
ranging from elementary particle experiments, such as 
resonance analyzed with the Breit--Wigner formula\cite{perkins} to
resonant electrical circuits\cite{purcell},
it has recently been actively investigated in the context of
biological information processing as an application
of stochastic resonance (see, e.g.,\cite{longtin,moss,collins,bulsara}).
In these studies, noise, which is normally considered 
an obstacle to information processing, is treated as an enhancer of
information processing through its resonance with external
signals.
It is known that delay, which is another element present and studied in 
biological phenomena and information processing (see, e.g.,\cite{mackey,milton1}), 
can cause a complex oscillatory behavior in an otherwise
simple, stable dynamical system. Analogous resonant phenomena
with respect to delay have also been noted and investigated.
In \cite{dr}, for example, the effect of delay in a lateral inhibition in a 
neural network is investigated both experimentally and theoretically.
It is found that delayed lateral inhibition can cause amplification 
of neural responses to sinusoidal stimuli in spite of the 
fact that, without delay, these inhibition generally attenuates
 such responses.

The main theme of this paper is presentation of a simple model 
which shows that delay and noise can have a resonance 
between themselves without an external periodic driving force.
Our model is a two state system whose dynamics is
governed by combinations of 
its state at some fixed interval in the past
and noise with a certain width.
In the probability space, this model can be described as a stochastic 
binary element
 whose transition probability
depends on its state at some fixed interval in the past.
Thus, in this description 
the model can be considered as a previously studied 
``delayed random walk''\cite{ohira1}
 except that it can only take two states.
With a fixed delay, we show analytically and numerically 
that such a model has a 
residence time histogram which shows a peak of maximum height
with appropriately chosen transition probability, i.e.,
noise and delay ``tuned'' together exhibit a resonance.
From the point of view of stochastic resonance,
this is a new type and one of the simplest
models which is analytically tractable.
We conclude the report with a discussion of possible utilization
of such a resonant behavior with noise and delay. 

Before presenting our model, we briefly discuss a
numerical study on a simple one dimensional system to
illustrate the physics motivating proposal of the model. 
Let us consider a simple one dimensional map dynamics with 
noise $\xi$ and delay $\tau$, which
is formally given by
\begin{equation}
z(t+1)= \tanh[\beta(z(t-\tau) - \theta)] + \xi_L 
\label{map}
\end{equation}
$\beta$ and $\theta$ are parameters and $\xi_L$ has the following
probability distribution.
\begin{eqnarray}
P(\xi=u) &=& {1 \over {2 L}}\quad (-L \leq u \leq L) \nonumber\\
         &=& 0 \quad (u < -L, u > L)
\end{eqnarray}
In words, $\xi_L$ is a time uncorrelated uniformly distributed noise 
taking the range $(-L, L)$. 
This map can be considered as a discrete time correspondence of the
following differential equation model.
\begin{eqnarray}
{dz \over dt} & = & -{{\partial \over {\partial z}}}V(z) + \xi_L \nonumber\\
V(z) & = & {1 \over 2} z^2(t) - {1 \over \beta}\log[\cosh(\beta(z(t-\tau)-\theta))]
\label{diff}
\end{eqnarray}
The shape of the asymmetric potential $V(z)$ with no delay is shown in Fig 1(A).
We have numerically simulated the map (\ref{map}) with various noise width and delay and found that we can have a regular spiking 
behavior as shown in Fig 1(C) for tuned noise
width and delay. Also, we have observed that the signal to noise ratio of the
corresponding peak in the Fourier spectrum
goes through a maximum with varying noise width as is generally
found in a system showing a stochastic resonance\cite{note1}.
We can qualitatively argue that the delay 
alters the effective potential
into an oscillatory one just like that due to an external oscillating force
leading to a stochastic resonance with tuned noise width.
The analysis of the dynamics given 
by (\ref{map}) or (\ref{diff}), however, is a non--trivial
task. Our model is an approximate abstraction of this dynamics retaining
asymmetric stochastic transition and delay in order to
gain insight into the resonant behavior between noise and delay.
\begin{figure}
\dofig{8.5cm}{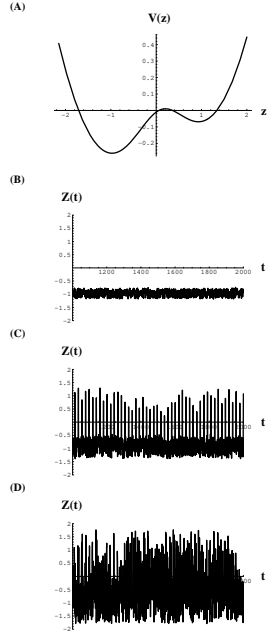}
\caption{
(A) The shape of the asymmetric potential for $\beta = 2.0$ 
and $\theta = 0.1$. Also the typical dynamics of $Z(t)$ from the
map model as we change noise width $L$. The values of $L$ are (B) $L = 0.2$, 
(C) $L = 0.4$, (D) $L =0.8$.
The data is taken with $\tau = 20$, $\beta = 2.0$, $\theta = 0.1$ and  
the initial condition $Z(t) = 0.0$ for $t \in [-\tau, 0]$. 
The plots are shown between $t=1000$ to $2000$.
} 
\label{fig1} 
\end{figure}

Let us now describe our model in detail.
The state of the system $X(t)$ at time
step $t$ can take either $-1$ or $1$. 
With the same noise $\xi_L$, we can define our model formally.
\begin{eqnarray}
X(t+1) &=& \theta[f(X(t-\tau)) + \xi_L], \nonumber\\
f(n) &=& {1 \over {2}}((a+b) + n(a-b)), \nonumber\\
\theta[n] &=& 1\quad (0 \leq n), \quad\quad -1\quad (0 > n),
\end{eqnarray}
where $a$ and $b$ are parameters such that $|a|\leq L$ and $|b|\leq L$, 
and $\tau$ is the delay. In relation to the map (\ref{map}), 
this model is an approximate discretization of space into
two states with $a$ and $b$ controlling
the bias of transition (reflecting the two different barrier heights from
the two stationary points of the potential), depending on the state of $X$ at $\tau$
steps before.

This model can be described in the probability space
more concisely as shown in Figure 2. 
\begin{figure}[h]
\dofig{2.4cm}{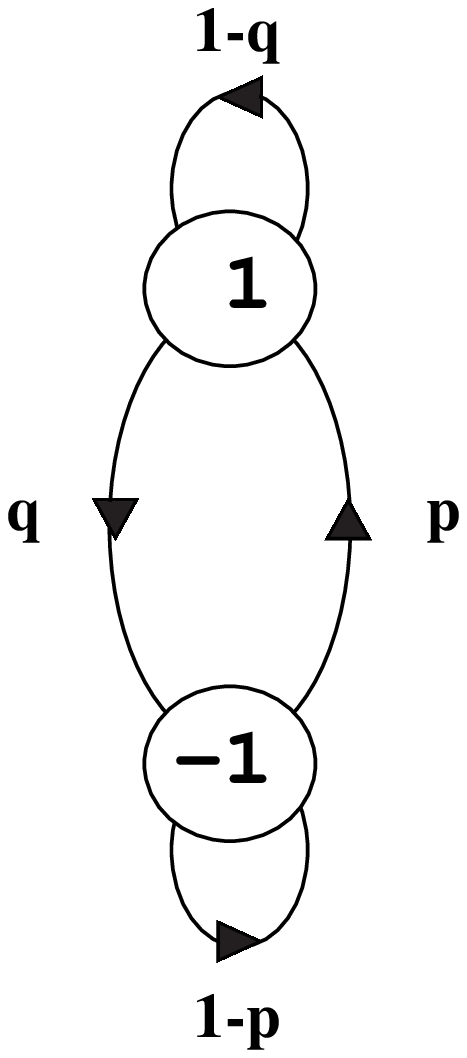}
\caption{
Schematic view of the stochastic model dynamics. 
} 
\label{fig2} 
\end{figure}
The formal definition is given as follows:
\begin{eqnarray}
P(1, t+1) &=& p,\quad \quad X(t-\tau) = -1,\nonumber\\
               &=& 1-q, \quad X(t-\tau) = 1,\nonumber\\
P(-1, t+1) &=& q, \quad \quad X(t-\tau) = 1,\nonumber\\
                &=& 1-p, \quad X(t-\tau) = -1,\nonumber\\
p &=& {1 \over {2}}(1+{b\over {L}}), \nonumber\\
q &=& {1 \over {2}}(1-{a\over {L}}),
\end{eqnarray}
where $P(s, t)$ is a probability that $X(t)=s$.  
Hence, the transition probability
of the model depends on its state at $\tau$ steps past and is a 
special case of delayed random walks\cite{ohira1}.

We first investigate the model numerically and observe that
a qualitatively similar feature to those shown in Figure 1 
appears.
We randomly
generate $X(t)$ for the interval $t=(-\tau, 0)$. 
Simulations are performed in which parameters 
are varied and $X(t)$ is recorded up to $10^6$ steps. 
From the trajectory $X(t)$, we construct a residence
time histogram $h(u)$ for the system to be in the state $-1$ for $u$
consecutive steps.
Some examples of histograms and corresponding $X(t)$
are shown in Figure 3 ($q = 1-q = 0.5$, $\tau=10$).
We note that with $p << 0.5$, as in Figure 3(A), the model
has a tendency to switch or spike to $X=1$ state after
the time step interval of $\tau$. But the spike trains do not
last long and result in a small peak in the histogram.
For the case of Figure 3(C) where $p$ is closer to $0.5$,
we observe less regular transitions and the peak height is 
again small. With appropriate $p$ as in Figure 3(B), 
spikes tend to appear at interval $\tau$ more frequently,
resulting in higher peaks in the histogram. 
\begin{figure}[tbh]
\dofig{9cm}{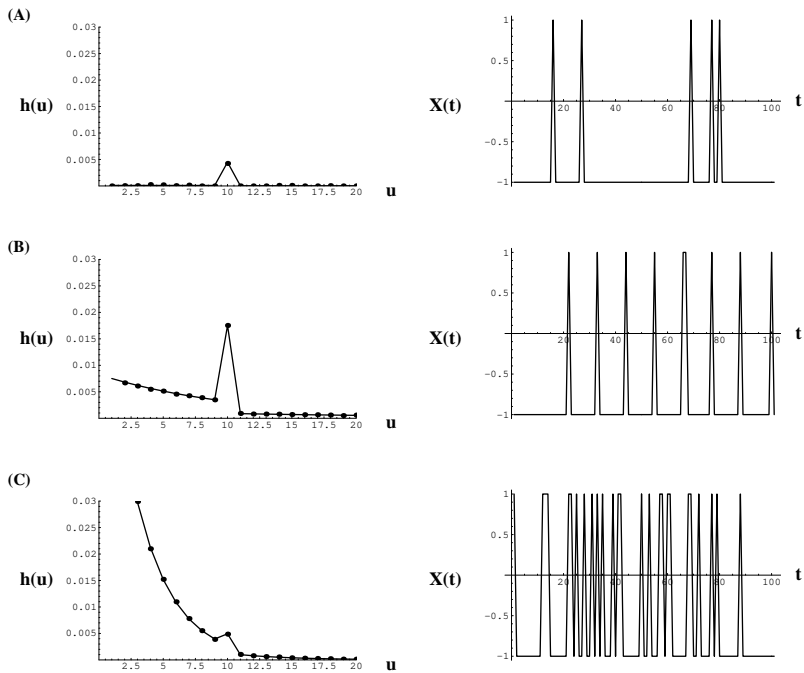}
\caption{
Residence time histogram and dynamics of $X(t)$ as we change $p$.
The values of $p$ are (A) $p=0.005$, (B) $p=0.05$, (C) $p=0.2$.
The solid line in the histogram is from the analytical expression
given in equations (8,9,10)
} 
\label{fig3} 
\end{figure}

This change of peak height in histograms which reaches maximum at an appropriate
noise level is one way to characterize stochastic resonance.
Choosing an appropriate $p$ is equivalent to "tuning" noise
width $L$ with other parameters appropriately fixed.
In this sense, our model shares a feature of the stochastic resonance.
It can be classified among models of stochastic resonance
without an external signal\cite{nosig}. The difference and the new point is 
 the use of delay as a source of its oscillatory dynamics. With this
characteristic, it could be termed as stochastic resonance with
delayed dynamics or, equivalently, a resonance between noise and delay.

In order to make this point clearer, let us treat the 
model analytically. 
The first observation to make with the model is that 
given $\tau$, it consists of statistically independent
$\tau + 1$ Markov chains. Each Markov chain has its
state appearing at every $\tau+1$ interval.
With this property of the model, we label time step $t$
by the two integers $s$ and $k$ as follows
\begin{equation}
t = s(\tau+1) + k, \quad (0 \leq s, 0 \leq k \leq \tau)
\end{equation}
Let $P_{\pm}(t) \equiv P_{\pm}(s, k)$ be the probability for the
state to be in the $\pm 1$ state at time $t$ or $(s, k)$.
Then, it can be derived that
\begin{eqnarray}
P_{+}(s,k)&=& \alpha(1-\gamma^s) + \gamma^s P_{+}(s=0,k),\nonumber\\
P_{-}(s,k)&=& \beta(1-\gamma^s) + \gamma^s P_{+}(s=0,k),\nonumber\\
\alpha &=& {p \over p+q},\nonumber\\
\beta &=& {q \over p+q},\nonumber\\
\gamma &=& 1 - (p+q).
\end{eqnarray}
In the steady state, we have $P_{+}(s\rightarrow \infty, k) \equiv P_{+} = \alpha$
and $P_{-}(s\rightarrow \infty, k) \equiv P_{-} = \beta$. The steady state residence time
histogram can be obtained by computing the following quantity,
$h(u) \equiv P(+; -, u; +)$, which is the probability that the 
system takes consecutive $-1$ state $u$ times between two $+1$ states. 
With the definition of the model and the property of statistical
independence between Markov chains in the sequence, the following
expression can be derived:
\begin{eqnarray}
P(+; -, u; +) & = & P_{+}(P_{-})^{u}P_{+} = 
(\beta)^{u}(\alpha)^{2} \quad (1 \leq u < \tau)\\
              & = & P_{+}(P_{-})^{\tau}(1-q) = 
(\beta)^{\tau}(\alpha)(1-q) \quad (u = \tau)\label{eqtau}\\
              & = & P_{+}(P_{-})^{\tau}(q)(1-p)^{u-\tau}(p) = 
(\beta)^{u}(p)^{2} \quad (u > \tau)
\end{eqnarray}
With appropriate normalization, this expression can reflect 
the shape of the histogram obtained by numerical simulations 
as shown in Figure 3.
Also, by differentiating equation (\ref{eqtau}) with respect to $p$, 
we can derive the resonant condition for the peak to reach maximum height as 
\begin{equation}
q = p\tau 
\label{taures}
\end{equation}
or, equivalently,
\begin{equation}
L-a = (L+b)\tau.
\label{taures2}
\end{equation}
Figure 3 shows how the maximum height changes
with $p$ and $\tau$. We see that peak maximum is reached
by choosing parameters according to equation (\ref{taures}).
Also, by changing $q$, we can control the width and height of
graphs in Figure 4. An example is shown in Figure 5.
We note that this analysis for the histogram is exact in the stationary limit, 
which is another feature of the model.
\begin{figure}
\dofig{8cm}{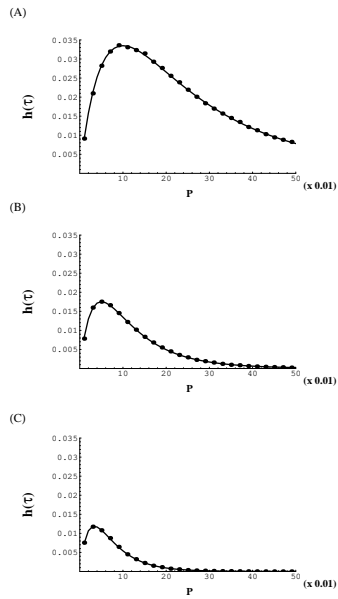}
\caption{
A plot of peak height by varying $p$
The solid
line is from equation (9).
} 
\label{fig4} 
\end{figure}
\begin{figure}
\dofig{4cm}{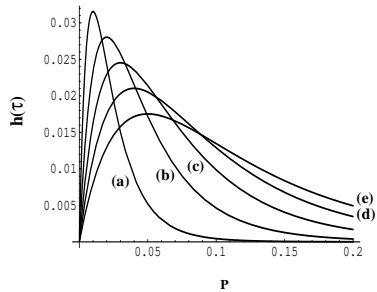}
\caption{
Resonance curves with various $q$ from equation (9).
The parameters are $\tau=10$, and $q=$ (a)0.1, (b)0.2, (c)0.3, (d)0.4, (e)0.5.} 
\label{fig5} 
\end{figure}

We have proposed here a very simple model which
nonetheless illustrates resonance behavior between noise and
delay both numerically and analytically. 
By 
relating $a$ and $b$ to the membrane threshold, the model
could be used as a very simplified and abstracted model
of spiking neuron with delayed self--feedback \cite{foss},
and could be developed for a model of pulse coupled 
neurons with delay\cite{ernst}.
Also, with the analytically tractable resonant
characteristics of the model described in this paper,
we could possibly seek an application of this delayed
stochastic binary element for information processing.
For example, one may be able to code temporal information
stochastically, or an encryption scheme
might be developed by extending this model\cite{ohira2}.
Exploration of such applications as well as 
extension of the model into many body systems are
the focus of our current research.


\begin{thebibliography}{99}
\bibitem{note1}
Details of the study of this model will be
discussed elsewhere.
\end{thebibliography}
\end{document}